\documentclass[aps,prc,twocolumn,showkeys,showpacs]{revtex4}
\usepackage{amssymb}% Include figure files
\usepackage{graphicx}% Include figure files
\usepackage{dcolumn}% Align table columns on decimal point
\usepackage{bm}% bold math

\newcommand{\req}[1]{(\ref{#1})}
\newcommand{\bel}[1]{\begin{equation}\label{#1}}
\newcommand{\belar}[1]{\begin{eqnarray}\label{#1}}

\def\pr{\prime}
\def\dpr{\prime\prime}

\begin{document}
\title{On the Scission Point Configuration of Fisioning Nuclei}
\author{F. A. Ivanyuk}
\affiliation{Institute for Nuclear Research, Prospect Nauki 47,
03028 Kiev, Ukraine}
\email{ivanyuk@kinr.kiev.ua}
\date{\today}
\begin{abstract}
The scission of a nucleus into two fragments is at present the
least understood part of the fission process, though the most
important for the formation of the observables.
To investigate the potential energy landscape at the largest
possible deformations, i.e. at the scission point (line,
hypersurface), the Strutinsky's optimal shape approach is applied.

For the accurate description of the mass-asymmetric nuclear shape
at the scission point, it turned out necessary to construct an
interpolation between the two sets of constraints for the
elongation and mass asymmetry which are applied successfully at
small deformations (quadrupole and octupole moments) and for
separated fragments (the distance between the centers of mass and
the difference of fragments masses).
In addition, a constraint on the neck radius was added, what
makes it possible to introduce the so called super-short and
super-long shapes at the scission point and to consider the
contributions to the observable data from different fission
modes. The calculated results for the mass distribution of the
fission fragment  and the Coulomb repulsion energy
"immediately after scission" are in a reasonable agreement with
experimental data.
\end{abstract}


\maketitle

%%%%%%%%%%%%%%%%%%%%%%%%%%%%%%%%%%%%%%%%%%%%%%%%%%%%%%%%%%%%%%%%%%%%
\section{Introduction}
\label{sec: intro}
%%%%%%%%%%%%%%%%%%%%%%%%%%%%%%%%%%%%%%%%%%%%%%%%%%%%%%%%%%%%%%%%%%%%

The shape of a nuclear surface is a basic notion in many
theoretical models of nuclear structure and reactions. A good
choice of the shape degrees of freedom reduces substantially the
computation time and is often, especially for the description of
fission process or fusion-fission reactions, a key to the success
of the theory.

In past a lot of shape parameterizations were proposed and used.
One class of shapes relies on the expansion in a complete set of
functions like the expansion of the radius vector $R(\theta)$
\cite{cohswi} or the profile function squared $\rho^2(z)$
\cite{koonin} in Legendre polynomials. In the parametrization
\cite{pash71} the deviation of the shape from the basic Cassini
ovals is also expanded in Legendre polynomials. Another
possibility is given by the introduction of a restricted number
of deformation parameters, like in
%%three smoothly joined quadratic surfaces parametrization \cite{nix},
the parametrization of three smoothly joined quadratic surfaces
\cite{nix}, the two center shell model \cite{marhun}, the
Funny-Hills parametrization \cite{brdapa} or modified
Funny-Hills parametrization \cite{pombar}.

All these shape parametrizations are restricted to a certain class
of shapes. In all these cases the question arises whether the
given class of shapes is complete enough
to represent the essential properties of the investigated process.

A method to introduce the shape of the nuclear surface
which does not rely on any shape parametrization was suggested by
V.Strutinsky already in \cite{stlapo,strjetp45}. In this method
one defines the profile function $\rho(z)$ of an axially
symmetric nucleus by the minimization of the liquid drop energy
with respect to the variation of $\rho(z)$ under additional
constraints which fix the volume and elongation of the drop.
However, due to numerical difficulties this method was not widely
used in the past.

Only recently \cite{fivan08} it turns out possibly to solve the
variational problem of \cite{stlapo} in a very broad region of
deformations ranging from a disk (even with a central depression)
to
two touching spheres. The fission barriers calculated by this
method \cite{ivapom2} were found to be in a reasonable agreement
with the experimental results.

In the present work the potential energy landscape is
investigated for larger deformations - at the scission point
(line, hypersurface). The scission of a nucleus into two
fragments is at present the least understood part of the fission
process, though the most important for the formation of the
observable data.

For the accurate description of the mass-asymmetric nuclear shape
at the scission point it turned out necessary to construct an
interpolation between the two sets of constraints for the
elongation and mass asymmetry which are applied successfully at
small deformations (quadrupole and octupole moments) and for
separated fragments (the distance between centers of mass and the
difference of fragments masses). In addition, a constraint on the
neck radius was added, what makes it possible to introduce the so
called super-short and super-long shapes at the scission point
and consider the contributions to the observable data from
different fission modes.

The paper is organized as follows. Section \ref{optimal} is a
short overview of the Strutinsky optimal shapes prescription. The
mass-asymmetric shapes are introduced in Sect. \ref{asymm}. The
scission shape and the shape of separated fragments "immediately
after scission"  are defined in Sect.
\ref{scission}-\ref{fragments}. In Sect. \ref{super} the
super-short and super-long shapes are introduced and the
calculated Coulomb repulsion energy of the fragments "immediately
after scission" is compared with the experimental total kinetic
energy of fission fragments of $^{236}U$. Sect. \ref{summa}
contains a short summary.
%%%%%%%%%%%%%%%%%%%%%%%%%%%%%%%%%%%%%%%%%%%%%%%%%%%%%%%%%%%%%%%%%%%%%%%%%%%%%%%%%%%%%%
\section{The optimal shapes of fissioning nuclei}
\label{optimal}
%%%%%%%%%%%%%%%%%%%%%%%%%%%%%%%%%%%%%%%%%%%%%%%%%%%%%%%%%%%%%%%%%%%%%%%%%%%%%%%%%%%%%%
The shape of an axially symmetric nucleus can be defined by
rotation of some profile function $\rho(z)$ around the $z$-axis.
It was suggested in \cite{stlapo} to define the profile function
looking for the minimum of the liquid-drop energy, $E_{\rm
LD}=E_{\rm surf} + E_{\rm Coul}$, under the constraint that the
volume $V$ and the elongation $R_{12}$ are fixed,
 \bel{variation}
 \frac{\delta}{\delta \rho}(E_{\rm LD} - \lambda_1 V -\lambda_2 R_{12}) = 0 \,,
\end{equation}
with
 \bel{r12}
 R_{12}=\frac{2\pi}{V}\int\limits_{z_1}^{z_2}\rho^2(z)\vert z\vert dz \,\,,
\quad V=\pi\int\limits_{z_1}^{z_2}\rho^2(z) dz \,\,.
\end{equation}
In \req{variation} $\lambda_1$ and $\lambda_2$ are the
corresponding Lagrange multipliers. The elongation parameter
$R_{12}$ was chosen by \cite{stlapo} to be the distance between
the centers of mass of the left and right parts of the nucleus,
%%%%%%%%%%%%%%%%%%%%%%%%%%%%%%%%%%%%%%%%%%%%%%%%%%%%%%%%%%%%%%%%%%%%%%%%%%%%%%%%%%%%
\begin{figure}[b]
\includegraphics[width=0.45\textwidth]{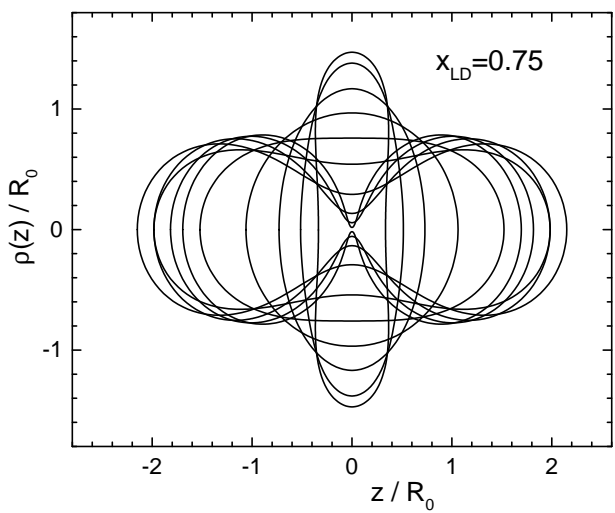}
\caption{The solutions of Eq. (\protect\ref{diffeq}) corresponding to
different values of Lagrange multiplier $\lambda_2$ which fixes
the elongation $R_{12}$}\label{fig1a}
\end{figure}
%%%%%%%%%%%%%%%%%%%%%%%%%%%%%%%%%%%%%%%%%%%%%%%%%%%%%%%%%%%%%%%%%%%%%%%%%%%%%%%%%%%%%%%%

The minimization of $E_{\rm LD}- \lambda_1 V -\lambda_2 R_{12}$
with respect to the profile function $\rho(z)$ leads to an
integro-differential equation for $\rho(z)$
 \bel{diffeq}
\rho\rho^{\prime\prime} = 1+(\rho^{\prime})^2-\rho[\lambda_1 +
\lambda_2 \vert z
 \vert
                  - 10x_{\rm LD}\Phi_S] [1+(\rho^{\prime})^2]^\frac{3}{2}\,.
\end{equation}
Here $\Phi_S\equiv\Phi(z, \rho(z))$ is the Coulomb potential at
the nuclear surface, and $x_{\rm LD}$ is the fissility parameter
of the liquid drop \cite{bohrwhee},
\begin{equation}\label{xldm}
x_{\rm LD}\equiv \frac{E_{\rm Coul}^{(0)}}{2E_{\rm surf}^{(0)}}
   =\frac{3}{10}\frac{Z^2e^2}{4\pi R_0^3\sigma}\approx \frac{Z^2}{49A}\,\,,
\end{equation}
where $\sigma$ is the surface tension coefficient.
In (\ref{xldm}) and everywhere below the index $^{(0)}$ refers to the spherical shape.

\begin{figure}[ht]
\includegraphics[width=0.45\textwidth]{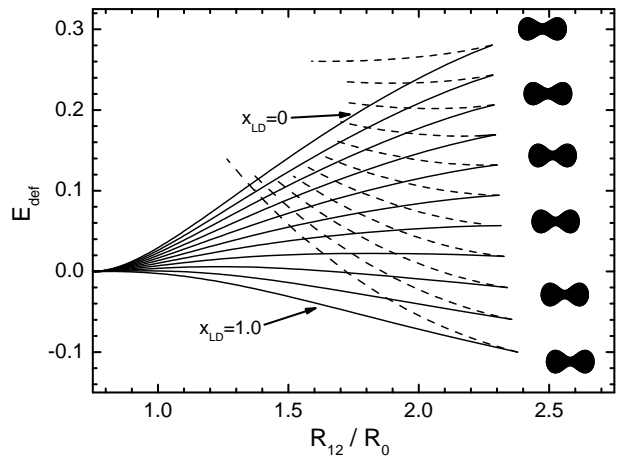}
\caption{Liquid-drop deformation energy
(\protect\ref{eldm}) as a function of the parameter $R_{12}$ for different
fissility parameters $x_{\rm LD}$ and the corresponding nuclear
shapes at scission.}\label{fig1b}
\end{figure}

By solving Eq. (\ref{diffeq}) one obtains the profile function
$\rho(z)$ for given $x_{LD}$ and $\lambda_2$ ($\lambda_1$ is
fixed by the volume conservation condition). The liquid drop
deformation energy $E_{\rm def}^{LD}=E_{\rm LD}-E_{\rm LD}^{(0)}$
(in units
 of the surface energy for
a spherical shape)
 \bel{eldm}
E_{def}\equiv E_{\rm def}^{LD}/{E_{\rm surf}^{(0)}}=B_{\rm surf}-1 + 2x_{\rm
LD}(B_{\rm Coul}-1)\,,
\end{equation}
calculated for the shapes shown in Fig.~\ref{fig1a}, is presented
in Fig.~\ref{fig1b}. In (\ref{eldm}) $B_{\rm Coul}\equiv {E_{\rm
Coul}}/{E_{\rm Coul}^{(0)}}$,  $B_{\rm surf}\equiv {E_{\rm
surf}}/{E_{\rm surf}^{(0)}}$.
%where an index $^{(0)}$ refers to the spherical shape.
%%%%%%%%%%%%%%%%%%%%%%%%%%%%%%%%%%%%%%%%%%%%%%%%%%%%%%%%%%%%%%%%%%%%%%%%%%%%%%%%%%%%
%%%%%%%%%%%%%%%%%%%%%%%%%%%%%%%%%%%%%%%%%%%%%%%%%%%%%%%%%%%%%%%%%%%%%%%%%%%%%%%%%%%%%%%%

One can see from Fig.~\ref{fig1b} that the elongation $R_{12}$ of
the shapes shown in these figures is limited by some maximal
value $R_{12}^{\rm sci}$. Above this deformation mono-nuclear
shapes do not exist.  This critical deformation was interpreted in
\cite{stlapo} as the scission point. Note that, at scission the
{\it neck radius} is still rather large: the neck radius at the
critical deformation is approximately equal to $(0.25 - 0.30)R_0$
for a fissility parameter in the range $0.4\leq x_{LD}\leq 0.9$

Another peculiarity of Fig.~\ref{fig1b} is the upper branch of
the deformation energy at large deformation. Along this branch
the neck of the drop becomes smaller and smaller until the shape
turns into two touching spheres. Both branches are
solutions of Eq.\req{diffeq}. It turns out, that the upper branch
of $E_{\rm def}^{LD}$ corresponds not to the minimum but to the
maximum of the energy. Thus, it represents the ridge of the
potential energy surface between the fission and fusion valleys.
%%%%%%%%%%%%%%%%%%%%%%%%%%%%%%%%%%%%%%%%%%%%%%%%%%%%%%%%%%%%%%%%%%%%%%%%%%%%%%%%%%%%%%%
\section{The mass-asymmetric shapes}
\label{asymm}
\begin{figure}[ht]
\includegraphics[width=0.45\textwidth]{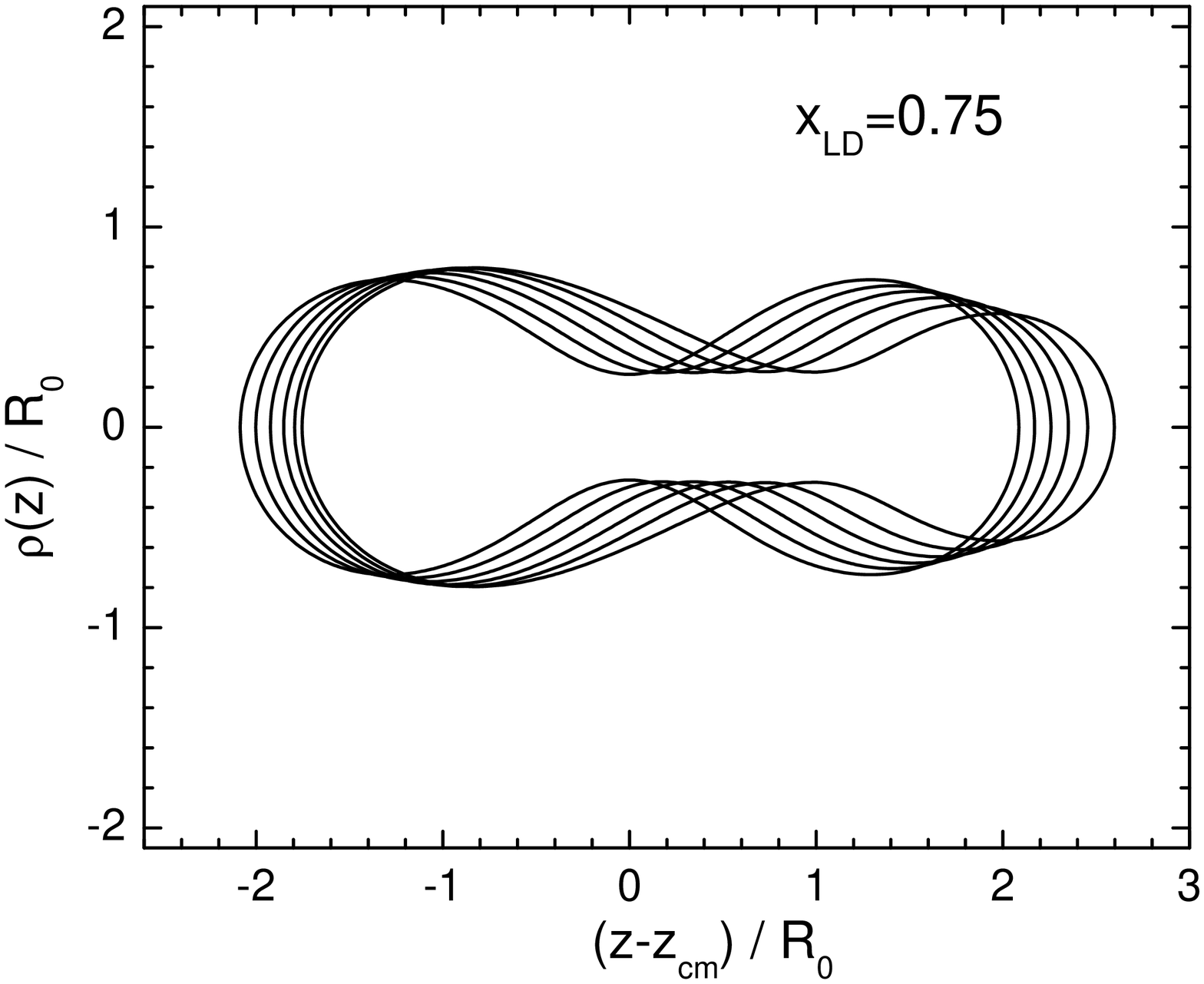}
\caption{The solutions of Eq. (\protect\ref{diffeq1}) at the maximal
elongation $R_{12}^{(sci.)}$ for a few values of the mass
asymmetry $\delta=0, 0.1, ... 0.5$}\label{fig2a}
\end{figure}

%%%%%%%%%%%%%%%%%%%%%%%%%%%%%%%%%%%%%%%%%%%%%%%%%%%%%%%%%%%%%%%%%%%%%%%%%%%%%%%%%%%%%%%
The optimal shape approach of \cite{stlapo} can be generalized to
mass-asymmetric shapes. For this aim one has to include into
Eq.\req{variation} one more constraint fixing the mass asymmetry
$\delta$ of the drop,
\bel{var1}
 \frac{\delta}{\delta \rho}(E_{\rm LD} - \lambda_1 V -\lambda_2 R_{12}-\lambda_3 \delta) = 0 \,.
\end{equation}
The mass asymmetry $\delta$ is commonly defined by the difference
of masses $M_L$ and $M_R$ to the left and right
of some point $z^*$,
\bel{delta}
\delta\equiv\frac{M_L-M_R}{M_L+M_R}=\frac{\pi}{V}\int \text{Sign}(z-z^*)\rho^2(z) dz\,.
\end{equation}
In case that the drop has a neck,
$z^*$ coincides with the position of the neck, $z^*=z_n$. By $z_n$ we mean here the point where $\rho(z)$ has a minimum.
For the pear-like shape the neck does not exist and $z^*$ could be defined in a different way, see \cite{fivan08}.
In the present work we are interested in the scission point
configuration for which the neck is well defined. Then the
Euler-Lagrange equation for the variational problem \req{var1}
has the form \belar{diffeq1}
\rho\rho^{\dpr}&=&1+(\rho^{\pr})^2-\rho\{\lambda_1 V + \lambda_2
\vert z-z_n \vert \\
&+&
\lambda_3 \text{Sign} (z-z_n)+ 10x_{LD}\Phi_S(z)\}[1+(\rho^{\pr})^2]^{3/2}\,.\nonumber
\end{eqnarray}
Eq.\req{diffeq1} can be solved in the same way as Eq.\req{diffeq}.
Some examples of the  shapes at the scission point
$R_{12}^{(sci.)}$ (maximal possible value of $R_{12}$) for few
values of the mass asymmetry are shown in Fig.~\ref{fig2a}.

%%%%%%%%%%%%%%%%%%%%%%%%%%%%%%%%%%%%%%%%%%%%%%%%%%%%%%%%%%%%%%%%%%%%%%%%%%%%%%%%%%%%
%%%%%%%%%%%%%%%%%%%%%%%%%%%%%%%%%%%%%%%%%%%%%%%%%%%%%%%%%%%%%%%%%%%%%%%%%%%%%%%%%%%%%%%%
The advantage of the variational problem in the form \req{var1}
is that the constraints for the elongation and the mass asymmetry
have a clear physical meaning. These are the distance $R_{12}$
between the centers of mass of right and left part of the drop and
the mass asymmetry $\delta$ of the drop.

The disadvantage is that due to the simplicity of the
restrictions on $R_{12}$ and $\delta$,
equation \req{diffeq1} contains some unphysical effects. Namely,
because
$\text{Sign}(z-z_n)$ is a discontinuous
function of $z$, the second order derivative $\rho^{\dpr}(z)$,
defined by Eq.\req{diffeq1}, and, consequently, the curvature of
the surface, is
discontinuous at $z=z_n$, what should not take place for a
liquid drop. Besides, for a pear-like shape the neck does not
exists and it is not so clear how one could define $z*$
in this case. Some possibility to define $z*$ as the place of the
largest curvature of the surface was suggested in \cite{fivan08}.

Besides $R_{12}$ and $\delta$
one could try another popular pair of constraints which are
often used in the constrained Hartee-Fock calculations,
namely, the quadrupole and octupole moments, \bel{q2q3}
Q_2=\frac{2}{V}\int dV r^2 P_2(\cos\theta)\,,
Q_3=\frac{1}{V}\int dV r^3 P_3(\cos\theta)\,.
\end{equation}
The moments $Q_2$ and $Q_3$ can be defined independently of
whether the neck exists or not. The use of
quadrupole and octupole moments as
constraints \bel{var2}
\frac{\delta}{\delta \rho}(E_{\rm LD} - \lambda_1 V -\lambda_2 Q_{2}-\lambda_3 Q_3) = 0 \,
\end{equation}
leads to the following Euler-Lagrange equation
\bel{diffeq2}
\rho\rho^{\dpr}=1+(\rho^{\pr})^2-\rho {\cal H}(z)[1+(\rho^{\pr})^2]^{\frac{3}{2}}\,,
\end{equation}
with
\belar{calhz}
{\cal H}(z)\equiv \lambda_1 V + \lambda_2 \left
[(z-z_{cm})^2- \frac{\rho^2(z)}{2}\right ]+  \\
\lambda_3 (z-z_{cm})\left [(z-z_{cm})^2-\frac{3\rho^2(z)}{2}\right ] +
10x_{LD}\Phi_S .\nonumber
\end{eqnarray}

\begin{figure}[ht]
\includegraphics[width=0.45\textwidth]{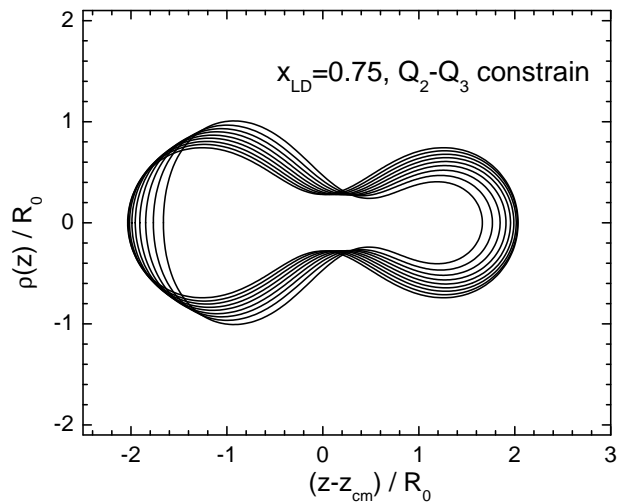}
\caption{The solutions of Eq.
(\protect\ref{diffeq2}) at the maximal elongation $R_{12}^{(sci.)}$ for
few values of the mass asymmetry $\delta=0, 0.1, ...
0.8$}\label{fig2b}
\end{figure}

Eq.\req{diffeq2} can be solved in the same way as
Eq.\req{diffeq1}. The examples of the scission point shapes for
few value of the mass asymmetry are shown in Fig.~\ref{fig2b}.
%%%%%%%%%%%%%%%%%%%%%%%%%%%%%%%%%%%%%%%%%%%%%%%%%%%%%%%%%%%%%%%%%%%%%%%%%%%%%%%%%%%%
%%%%%%%%%%%%%%%%%%%%%%%%%%%%%%%%%%%%%%%%%%%%%%%%%%%%%%%%%%%%%%%%%%%%%%%%%%%%%%%%%%%%%%%%

Comparing the shapes at the scission point calculated  by Eq.
\req{diffeq1} and Eqs. \req{diffeq2}-\req{calhz} for the same mass-asymmetry
one can see that these two sets of shapes are rather different.
The $Q_2, Q_3$ restrictions lead to
scission shapes which are considerably "shorter" as compared with
the scission shapes defined with $R_{12}, \delta $ restrictions.

\begin{figure}[ht]
\includegraphics[width=0.45\textwidth]{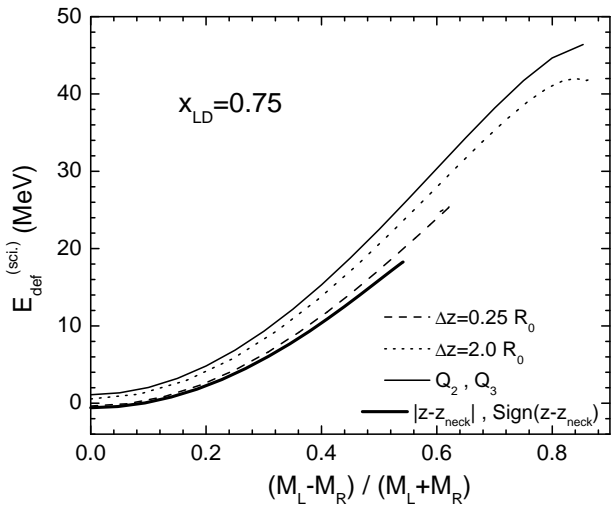}
\caption{Liquid drop energy (\protect\ref{eldm}) at the maximal elongation
$R_{12}^{(sci.)}$  calculated with solutions of Eq.\protect\req{diffeq1}
(thick solid line), Eq.\protect\req{diffeq2} (thin solid line) and
Eq.\protect\req{diffeq4} (dash and dot lines) as function of the mass
asymmetry} \label{fig3a}
\end{figure}

The comparison of the energies at the scission point is shown in
Fig.~\ref{fig3a}. Due to the smaller Coulomb repulsion energy for
more elongated
shapes the scission point energy calculated with the profile
function \req{diffeq1} is by $(2\div 5)$ MeV lower than that
calculated with the profile function \req{diffeq2}-\req{calhz}.

The expansion in multipole moments is an expansion in the
complete set of orthogonal functions. At small deformation only
few lower moments are important. At
large deformation, especially at the scission point, one can not
characterize the optimal shape of the surface by the quadrupole
and octupole moments
alone. Higher multipole moments should
then also be taken
into account.
%%%%%%%%%%%%%%%%%%%%%%%%%%%%%%%%%%%%%%%%%%%%%%%%%%%%%%%%%%%%%%%%%%%%%%%%%%%%%%%%%%%%
%%%%%%%%%%%%%%%%%%%%%%%%%%%%%%%%%%%%%%%%%%%%%%%%%%%%%%%%%%%%%%%%%%%%%%%%%%%%%%%%%%%%%%%%

The optimal shapes defined with $R_{12}, \delta$ constraints
describe well the separated or touching drops. For the shape with
a neck one should try to define a constraint which would be an
interpolation between $Q_2, Q_3$ and $R_{12}, \delta$ constraints.
Some hint how
this can be achieved, one can get looking at the curvature of the
surface calculated with both constraints.
At each point of the surface one can define the local curvature
$H(z)$,
\bel{hz}
  H(z)=\frac{1}{2}\left(\frac{1}{R_1}+\frac{1}{R_2}\right)\,,
\end{equation}
where $R_1$ and $R_2$ are the local principal radii of curvature.
In the case of axially symmetric shapes the radii $R_1$ and $R_2$
can be expressed in terms of the profile function $\rho(z)$,
\bel{r1r2}
  R_1=\rho(z)\sqrt{1+(\rho^{\prime})^2},
\quad R_2=-[1+(\rho^{\prime})^2]^\frac{3}{2}/\rho^{\prime\prime},
\end{equation}
Inserting \req{r1r2} into \req{hz} and solving this equation
with respect to $\rho^{\dpr}(z)$ one gets the following relation
between the profile function $\rho(z)$ and the local curvature
$H(z)$,
\bel{diffeq3} \rho\rho^{\prime\prime} =
1+(\rho^{\prime})^2-2 \rho H(z)
[1+(\rho^{\prime})^2]^\frac{3}{2}\,.
\end{equation}
Comparing Eq.\req{diffeq3} and Eq.\req{diffeq2} one sees that the
expression in curly brackets in \req{diffeq1} or \req{diffeq2} is
just twice the local curvature of the surface.

\begin{figure}[ht]
\includegraphics[width=0.45\textwidth]{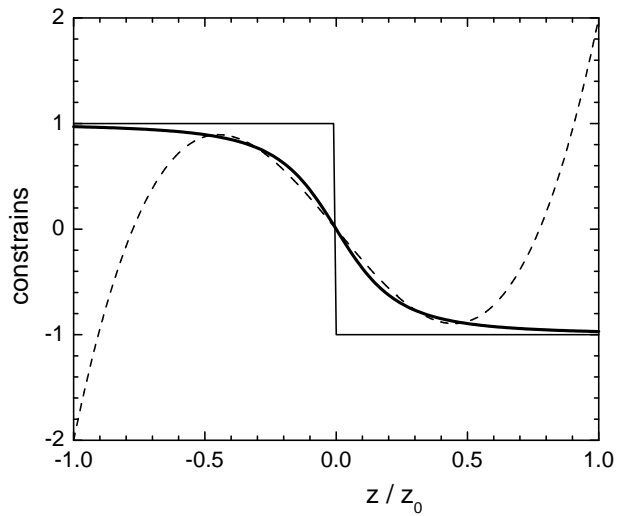}
\caption{Illustration of the constraints $\delta$ (solid),
$P_3(z/z_0)$ (dash) and smoothed Sign-function (thick
solid line)} \label{fig3b}
\end{figure}

The left-right asymmetric part of the curvature proportional to
$\text {Sign}\,z$ and $P_3(z/z_0)$ (the length of the drop along
the $z$-axes is equal to $2z_0$) is shown in Fig.~\ref{fig3b}. The
function $P_3(z/z_0)$ which appears in Eq.\req{var2} grows
rapidly at the tips of the drop (at $z\approx z_0$). At the tips
of the drop the surface of the heavy fragment becomes flat, the
surface of the light fragment becomes very deformed (elongated).
This is in contradiction with the expectation that due to Coulomb
repulsion the distant parts of the drop should be close to
spheroids.

The spherical shape has a constant curvature. I.e. for the
distant part of the drop the $\text {Sign}\,z$ constraint is more
meaningful than $Q_3$. It is also clear that at small $z$ (in the
neck region) the curvature should change smoothly between the
asymptotic values on the very left and on the very right.

%%%%%%%%%%%%%%%%%%%%%%%%%%%%%%%%%%%%%%%%%%%%%%%%%%%%%%%%%%%%%%%%%%%%%%%%%%%%%%%%%%%%
%%%%%%%%%%%%%%%%%%%%%%%%%%%%%%%%%%%%%%%%%%%%%%%%%%%%%%%%%%%%%%%%%%%%%%%%%%%%%%%%%%%%%%%%
These requirements can be fulfilled if instead of  $\text
{Sign}\,z$ one would introduce a smoothed $\text {Sign}$ -
function, say by the replacement
\bel{smooth} \vert z\vert
\Longrightarrow \sqrt{z^2 + (\Delta z)^2},\,\,\,\,
\text{Sign}\,z\Longrightarrow z/\sqrt{z^2 +
(\Delta z)^2}
\end{equation}
and use these smoothed quantities as the constraints for the
elongation and mass asymmetry.
In \req{smooth} we have introduced also a smoother $\vert z \vert$ - function which appears in the definition \req{r12} of $R_{12}$ constraint.
%%%%%%%%%%%%%%%%%%%%%%%%%%%%%%%%%%%%%%%%%%%%%%%%%%%%%%%%%%%%%%%%%%%%%%%%%%%%%
\section{The scission shapes}
\label{scission}
%%%%%%%%%%%%%%%%%%%%%%%%%%%%%%%%%%%%%%%%%%%%%%%%%%%%%%%%%%%%%%%%%%%%%%%%%%%%%%%%%%%%
The replacement \req{smooth} contains an additional parameter -
the smoothing width $\Delta z$. In principle, one can consider it
as an additional collective parameter which has to be taken into
account in the dynamical calculations. In the quasi-static limit
one could expect that the value of $\Delta z$ is close the
curvature  radius $R_2$ in the neck region.

\begin{figure}[ht]
\includegraphics[width=0.45\textwidth]{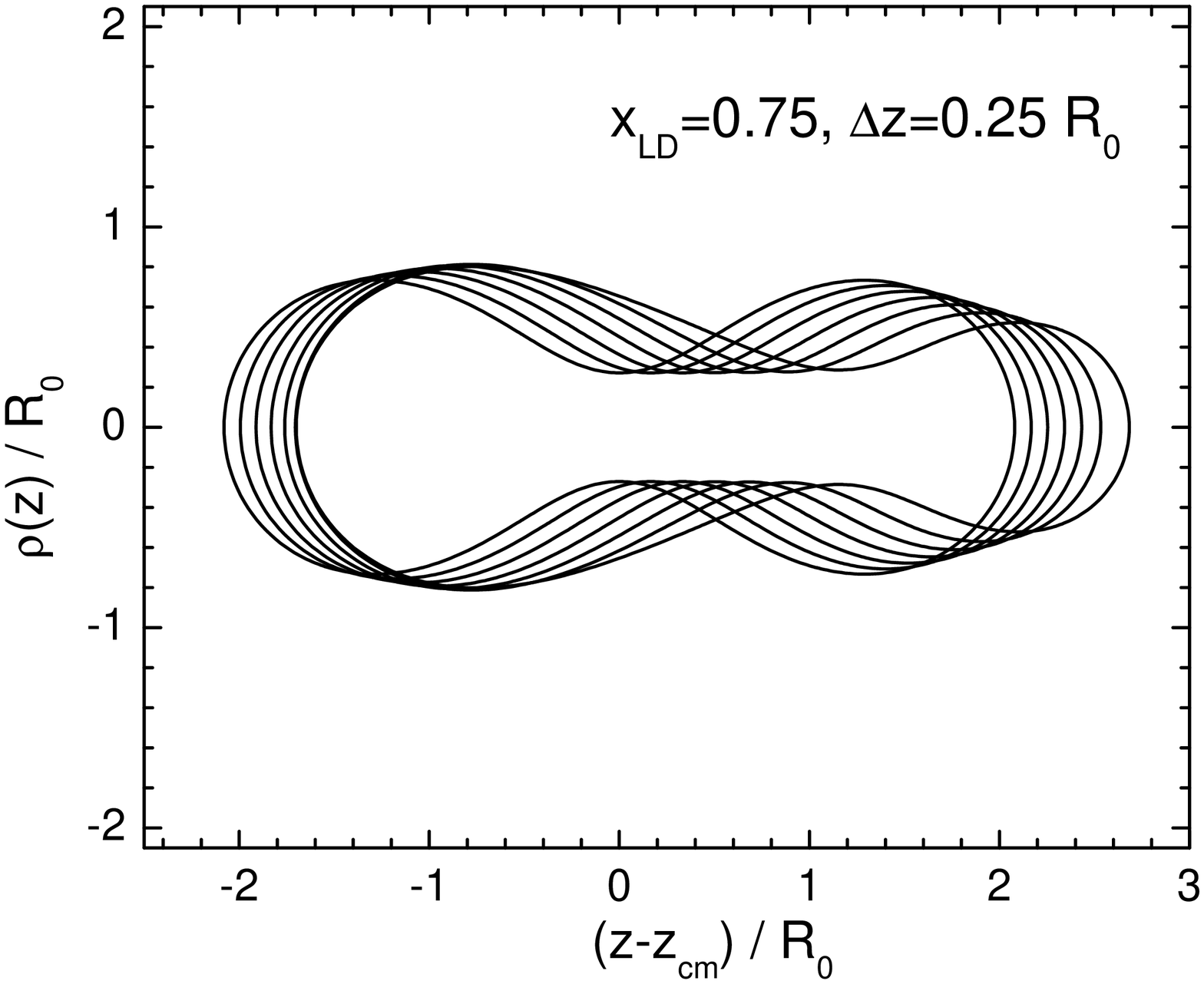}
\caption{The solutions of Eq. (\protect\ref{diffeq4}) ($\Delta z=0.25
R_0$) at the maximal elongation $R_{12}^{(sci.)}$ for few values
of the mass asymmetry $\delta=0, 0.1, ... 0.6$}\label{fig4a}
\end{figure}

\begin{figure}[ht]
\includegraphics[width=0.45\textwidth]{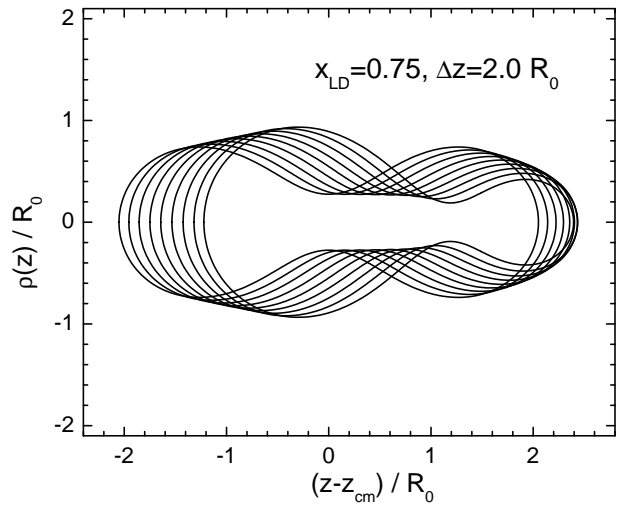}
\caption{The same as
in Fig.~\protect\ref{fig4a} calculated for $\Delta z=2.0 R_0$ and $\delta=0,
0.1, ... 0.8$}\label{fig4b}
\end{figure}

Replacing in \req{diffeq2} the $\vert z\vert$ and $\text{Sign}\,z$ by the smoothed quantities \req{smooth} one gets the following equation for $\rho(z)$
\belar{diffeq4}
\rho\rho^{\prime\prime} = 1+(\rho^{\prime})^2 - \rho \left\{\right.
\lambda_1 V + \lambda_2 \sqrt{(z-z_n)^2+(\Delta z)^2}+\nonumber \\
\left.\lambda_3\frac{(z-z_n)}{\sqrt{(z-z_n)^2+(\Delta z)^2}}
+10x_{LD}\Phi_S(z) \right\} [1+(\rho^{\prime})^2]^\frac{3}{2}.
\end{eqnarray}
In Figs.  \ref{fig4a}, \ref{fig4b} we show the optimal shape at the scission point
calculated with the constraint \req{smooth} for
two very different
%% THIS IS A SUGGESTION TRYING TO EXPLAIN WHY THESE VALUES HAVE BEEN CHOSEN
values of $\Delta z$, $\Delta z=0.25 R_0$ and $\Delta z=2.0 R_0$.
The first is approximately equal to the neck radius, the second -
to the half-length of the drop in $z$-direction at the scission
point.

%%%%%%%%%%%%%%%%%%%%%%%%%%%%%%%%%%%%%%%%%%%%%%%%%%%%%%%%%%%%%%%%%%%%%%%%%%%%%%%%%%%%
%%%%%%%%%%%%%%%%%%%%%%%%%%%%%%%%%%%%%%%%%%%%%%%%%%%%%%%%%%%%%%%%%%%%%%%%%%%%%%%%%%%%%%%%

The Figs.  \ref{fig4a}, \ref{fig4b} are similar to Figs.  \ref{fig2a}, \ref{fig2b}. The shapes
calculated with $\Delta z=0.25 R_0$ are rather close the shapes
calculated with $R_{12}, \delta$ constraints (in the limit $\Delta
z\to 0$ the profile functions shown in Fig.~\ref{fig2a} and
Fig.~\ref{fig4a} coincide). The energies calculated with $\Delta
z=0.25 R_0$ and $R_{12}, \delta$ constraints are also very close
to each another, see Fig.~\ref{fig3a}. With growing $\Delta z$ the
scission shapes are getting shorter like
those calculated with the $Q_2, Q_3$ constraints. However the
tips of the shapes calculated with $\Delta z=2.0 R_0$ are more
"spherical" as compared with
those calculated with $Q_2, Q_3$ constraints. The energies of the
shapes calculated with $\Delta z=2.0 R_0$ are by $(1\div 3)$ MeV
lower as compared  with
the ones calculated with $Q_2, Q_3$ constraints.

Taking into account the results shown in Fig.~\ref{fig3a},
the shapes calculated with $\Delta z=0.25 R_0$ seem more
preferable as the scission shapes. Besides, the total kinetic
energy of the fission fragments calculated with $\Delta z=0.25
R_0$ is in better agreement with the experimental data as
compared with the one calculated for $\Delta z=2.0 R_0$
(see the Fig.~\ref{fig7b} below).
So, in calculations below the shapes shown in Fig.~\ref{fig4a}
will be used as the scission shapes.
%%%%%%%%%%%%%%%%%%%%%%%%%%%%%%%%%%%%%%%%%%%%%%%%%%%%%%%%%%%%%%%%%%%%%%%%%%%%%%%%%%%%
%%%%%%%%%%%%%%%%%%%%%%%%%%%%%%%%%%%%%%%%%%%%%%%%%%%%%%%%%%%%%%%%%%%%%%%%%%%%%%%%%%%%%%%%%%%%%%%

The description of the fission process requires a solution of the
dynamical problem. The potential energy surface is
important,
but only one ingredient of the dynamical description. The
inertia, friction and diffusion tensors are also equally
important. Still, having only the potential energy surface at
ones disposal,
one could try to estimate some observable of the fission process.

Keeping in mind that the fission process is slow one could assume
that during the fission process the state of the fissioning
nucleus is close to
thermal equilibrium, i.e. in the quasistatic limit
the points $\{q_i\}$ on the deformation energy surface are
populated with the probability given by the Boltzman factor,
\bel{proba} P(q_i)=e^{-\frac{E(q_i)-E_0}{T_{coll}}}
\end{equation}
Here $T_{coll}$ is the temperature, and $E_0$ is a constant which
is not important for what follows.

Then, the normalized mass distribution of the fission fragments
will be defined by the deformation energy at maximal deformation
$R_{12}^{(sci.)}$ considered as a function of the mass asymmetry
$\delta_i$ for each pair of fragments,
\bel{yield} Y =\frac{e^{-E_{def}(R_{12}^{(sci.)},\,
\delta_i)/T_{coll}}}{\sum_ie^{-E_{def}(R_{12}^{(sci.)},\,
\delta_i)/T_{coll}}}\,.
\end{equation}
where $E_{def}(R_{12}^{(sci.)}\,,\delta_i)$ is the liquid drop
energy \req{eldm} plus the shell correction $E_{\text   shell}$.

\begin{figure}[htb]
\includegraphics[width=0.45\textwidth]{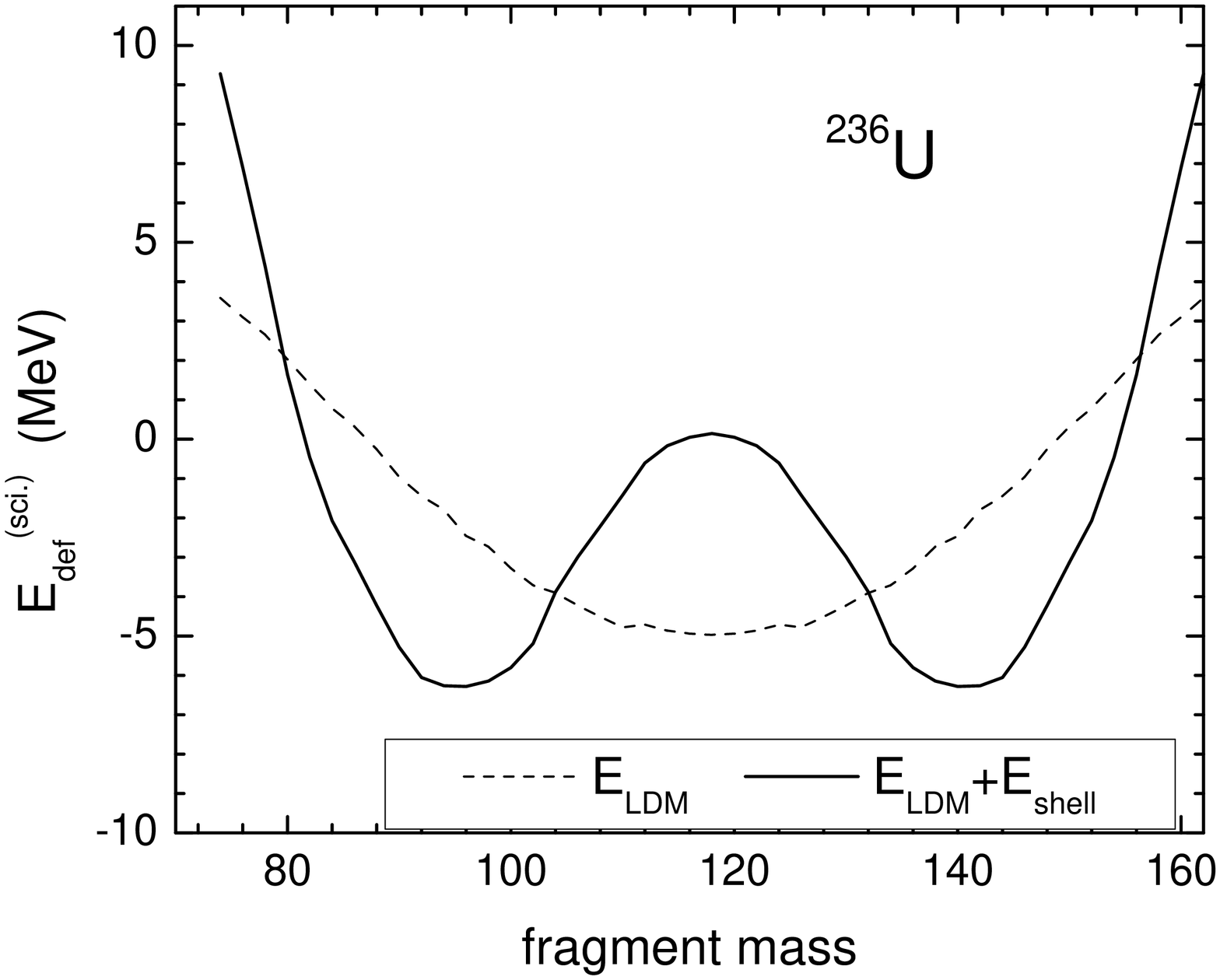}
\caption{The liquid drop (dash) and the total
$E_{tot}=E_{LDM}+E_{shell}$ deformation energy of $^{236}U$ along
the scission line (at maximal elongation $R_{12}^{(sci.)}$)
calculated with the solutions of Eq.\protect\req{diffeq4} for $\Delta
z=0.25 R_0$}\label{fig5a}
\end{figure}

To calculate the shell correction energy $E_{shell}$ we have
approximated the shapes shown in Fig.~\ref{fig4a} by the Cassini
ovaloids with three deformation parameters $\alpha, \alpha_1,
\alpha_2$, see \cite{pash71}, and for the shape given in terms of
Cassini ovaloids calculated the single-particle energies and the
shell correction by the code \cite{pash71}. The liquid drop
energy and the total energy including the shell correction for
the nucleus $^{236}U$ are shown in Fig.~\ref{fig5a}.
%%%%%%%%%%%%%%%%%%%%%%%%%%%%%%%%%%%%%%%%%%%%%%%%%%%%%%%%%%%%%%%%%%%%%%%%%
%%%%%%%%%%%%%%%%%%%%%%%%%%%%%%%%%%%%%%%%%%%%%%%%%%%%%%%%%%%%%%%%%%%%%%%%%%%%%%%%%%%%%%%%%%%%%%%%%%%%%%%%%%%%%%%%%%%%%%%%%%%%%%%%%%%%%%%%%%%%%%%%%%%%%%%%%%%%%%%%%
\begin{figure}[hb]
\includegraphics[width=0.45\textwidth]{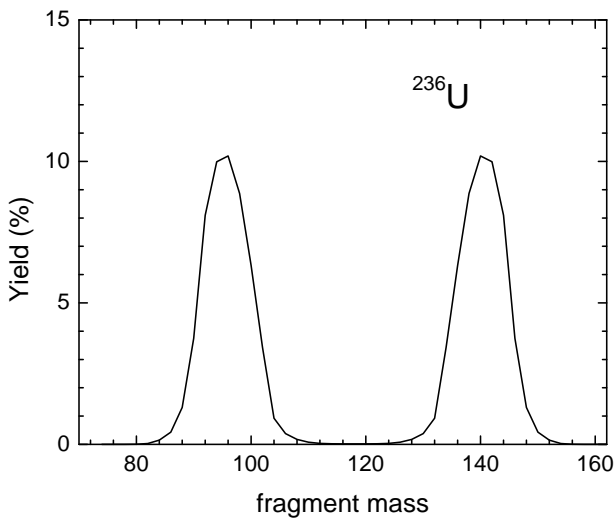}
\caption{The yield \protect\req{yield} of the fission fragments
of $^{236}U$ calculated with the deformation energy shown in Fig.~\protect\ref{fig5a}.}\label{fig5b}
\end{figure}
%%%%%%%%%%%%%%%%%%%%%%%%%%%%%%%%%%%%%%%%%%%%%%%%%%%%%%%%%%%%%%%%%%%%%%%%%%%%%%%%%%%%%%%%%%%%%%%%

The total
energy has a minimum at the fragment mass equal to 95 or 141.
Consequently the peaks of the mass distribution of the fission
fragment are located at these
fragments masses, see Fig.~\ref{fig5a}. The position of the peaks
of the mass distribution is in
good agreement with the well known experimental results
\cite{huizenga}. This agreement can be considered as a
confirmation that the scission point shape and energy are
calculated correctly.

The distribution \req{proba} is a basic assumption of the
scission-point model suggested by \cite{spm} and developed
further in \cite{spm2,spm3,spm4}, see also \cite{krapom}. In
this model the scision point configuration consists of two
coaxial spheroids with tip-to-tip distance $d$ and quadrupole
deformation parameters $\beta_L$ and $\beta_H$. There are two
temperatures in this model one for the population of the
single-particle levels, $T_{int}$,   and one for the collective
degrees of freedom, $T_{coll}$. The parameters $d$, $T_{int}$ and
$T_{coll}$ were fitted in \cite{spm} in order to reproduce the
experimental data. The $T_{coll}$ was found to be close to 1 Mev.
In the calculations shown in Fig.~\ref{fig5b} we used the same
value $T_{coll}$=1 MeV.

Note, that within the optimal shape approach the shape
configuration is not fitted to the experimental results but is
defined unambiguously from the minimal energy condition.
%%%%%%%%%%%%%%%%%%%%%%%%%%%%%%%%%%%%%%%%%%%
\section{The separated fragments}
\label{fragments}
%%%%%%%%%%%%%%%%%%%%%%%%%%%%%%%%%%%%%%%%%%%%%%%%%%%%%%%%%%%%%%%%%%%%%%%%%%%%%%%%%%%%%%%%%%%%%%%
Within the optimal-shape method one can also find the optimal
shape of separated fragments. For this aim one solves equation
\req{diffeq} with the initial conditions that correspond
to two spherical fragments at large enough distance $R_{12}$ from
each other. Making the distance $R_{12}$ smaller and smaller,
one can find out  how the shape of the fragments changes with the
distance between their centers of mass.
%%%%%%%%%%%%%%%%%%%%%%%%%%%%%%%%%%%%%%%%%%%%%%%%%%%%%%%%%%%%%%%%%%%%%%%%%%%%%%%%%%%%
\begin{figure}[ht]
\includegraphics[width=0.45\textwidth]{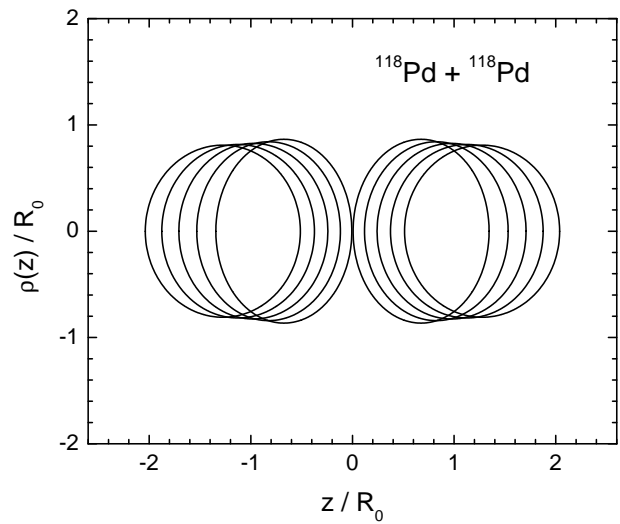}
\caption{The solutions of Eq. (\protect\ref{diffeq}) corresponding to
different values of the
Lagrange multiplier $\lambda_2$ which fixes the deformation
$R_{12}$}\label{fig6a}
\end{figure}
%%%%%%%%%%%%%%%%%%%%%%%%%%%%%%%%%%%%%%%%%%%%%%%%%%%%%%%%%%%%%%%%%%%%%%%%%%%%%%%%%%%%%%%%
The results of numerical calculations for the symmetric splitting
of $^{236}U$ are shown in Fig.~\ref{fig6a}. The shape of the
separated fragments is very close to oblate ellipsoids. The
octupole deformation is very small. Its contribution to the
deformation energy at the touching point of two $^{118}Pd$ nuclei
is of the order 0.5 Mev only.
The energy of separated fragments is shown by the dashed line in
Fig.~\ref{fig6b}. The lower solid curve and the dashed line
correspond respectively to the bottom of the fission and the
fusion valleys and the upper solid curve to the ridge between the
fusion and fission valleys.

The kinetic energy of the fission fragments is the kinetic energy
gained by the fragments due to the Coulomb repulsion after
separation plus the prescission kinetic energy. Within the
quasi-static picture one can calculate only the energy of
the
Coulomb repulsion ``immediately after
scission". At present it is not so clear how the scission process
proceeds.
For slow collective motion it is natural to assume that during
the neck rupture the elongation (the distance between centers of
mass of left and right parts of nucleus) does not change, like it
is shown by arrow in Fig.~\ref{fig6b}. The corresponding profile functions at $R_{12}=R_{12}^{(sci.)}$ for the compact system and separated fragments are show in Fig.~\ref{fig7a}.
%%%%%%%%%%%%%%%%%%%%%%%%%%%%%%%%%%%%%%%%%%%%%%%%%%%%%%%%%%%%%%%%%%%%%%%%%%%%%%%%%%%%
\begin{figure}[ht]
\includegraphics[width=0.45\textwidth]{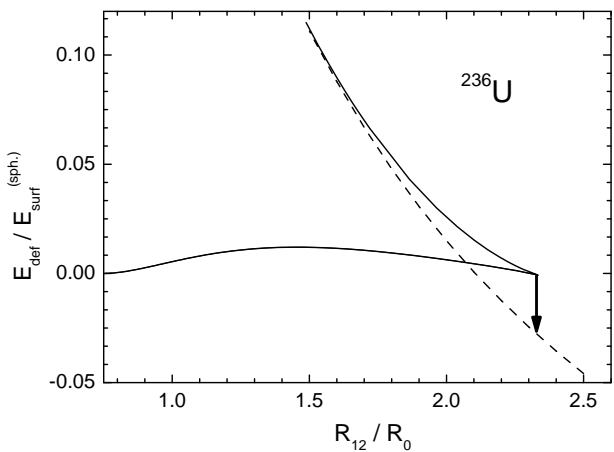}
\caption{Liquid-drop deformation energy (\protect\ref{eldm}) of
$^{236}U$ as a function of the parameter $R_{12}$ for  the
mono-nucleus (solid) and separated fragments (dash).
}\label{fig6b}
\end{figure}
%%%%%%%%%%%%%%%%%%%%%%%%%%%%%%%%%%%%%%%%%%%%%%%%%%%%%%%%%%%%%%%%%%%%%%%%%%%%%%%%%%%%%%%%

The Coulomb interaction energy of the fragments immediately after
scission shown in Fig.~\ref{fig7a} is easy to calculate. The only
additional parameter (besides the fissility parameter $x_{LD}$)
which appears in such calculation is the parameter $r_0$ of the
nuclear radius, $R_0=r_0 A^{1/3}$. In the
present work we used the value $r_0=1.225$ fm.
% as in previous works \cite{???,???}.
%%%%%%%%%%%%%%%%%%%%%%%%%%%%%%%%%%%%%%%%%%%%%%%%%%%%%%%%%%%%%%%%%%%%%%%%%%%%%%%%%%%%
\begin{figure}[htb]
\includegraphics[width=0.45\textwidth]{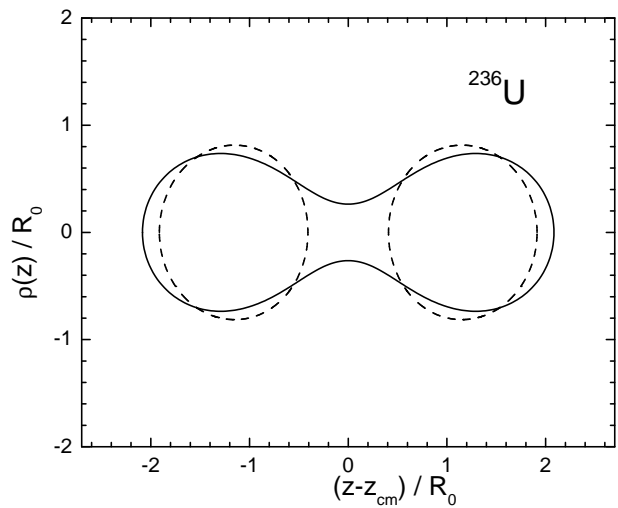}
\caption{Profile function, solution of Eq.\ (\protect\ref{diffeq4})
at the scission point $R_{12}^{(sci.)}$ for a mass-symmetric
deformation of $^{236}U$ (solid line), and profile function
(\protect\ref{diffeq}) of two separate fragments with a distance between
their centers of mass equal to $R_{12}^{(sci.)}$ (dashed lines)}\label{fig7a}
\end{figure}
%%%%%%%%%%%%%%%%%%%%%%%%%%%%%%%%%%%%%%%%%%%%%%%%%%%%%%%%%%%%%%%%%%%%%%%%%%%%%%%%%%%%%%%%

\begin{figure}[ht]
\includegraphics[width=0.45\textwidth]{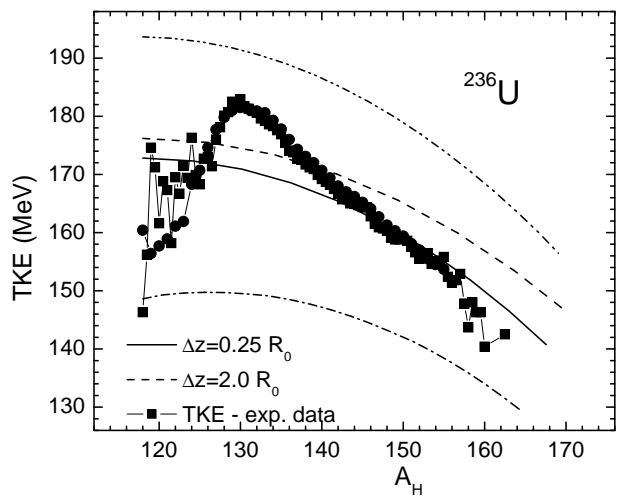}
\caption{Coulomb repulsion energy of two separated fragments at a distance $R_{12}=R_{12}^{(sci.)}$, where $R_{12}^{(sci.)}$ is the maximal elongation calculated with Eq.\req{diffeq4} with $\Delta z=0.25 R_0$ (solid line) and $\Delta z=2.0 R_0$ (dashed line). The experimental values of the total kinetic energy are taken from \protect\cite{mul,baba,zey} The dash-dot and dash-dot-dot lines show Coulomb repulsion energy calculated for super-long and super-short shapes (see Fig.~\protect\ref{fig8b})}\label{fig7b}
\end{figure}

The comparison of the Coulomb interaction energy of the fragments
immediately after scission with the experimental value of the total
kinetic energy for
the
nucleus $^{236}U$ is shown in Fig.~\ref{fig7b}. The solid and dashed lines are
calculated with $\Delta z=0.25 R_0$ (solid) and $\Delta z=2.0 R_0$ (dashed). One
sees that the more elongated scission ($\Delta z=0.25 R_0$)
shapes are in somewhat better agreements with the experimental
data.

The agreement is of
qualitative character only. For
a more accurate description on should take into account the
multimodal character of the fission of $^{236}U$. The optimal
shapes described above correspond to only one (standard) fission
mode.
%%%%%%%%%%%%%%%%%%%%%%%%%%%%%%%%%%%%%%%%%%%%%%%%%%%%%%%%%%%%%%%%%%%%%%%%%%%%%%%%%%%%
%%%%%%%%%%%%%%%%%%%%%%%%%%%%%%%%%%%%%%%%%%%%%%%%%%%%%%%%%%%%%%%%%%%%%%%%%%%%%%%%%%%%%%%%
%%%%%%%%%%%%%%%%%%%%%%%%%%%%%%%%%%%%%%%%%%%%%%%%%%%%%%%%%%%%%%%%%%%%%%%%%%%%%%%%%%%%%%%%
\section{Super-long and super-short scission shapes}
\label{super}
%%%%%%%%%%%%%%%%%%%%%%%%%%%%%%%%%%%%%%%%%%%%%%%%%%%%%%%%%%%%%%%%%%%%%%%%%%%%%%%%%%%%%%%%
The optimal shapes discussed in Sections 3-5 have the two degrees
of freedom - elongation and the mass asymmetry. The neck radius
for the given elongation and the mass asymmetry attains the "most
favored" value which results from the minimum of the potential
energy condition. In the dynamical calculations of the fission
process the neck radius is often considered as an independent
collective variable which can deviate from the one corresponding
to the bottom of the potential energy surface. Thus, it makes
sense to incorporate in the optimal shapes procedure the neck
radius as another independent degree of freedom.

In order to include one
additional degree of freedom in the optimal shapes procedure one
should add
another constraint fixing the neck radius. Usually, in various
shape parameterizations,
the neck radius is regulated by the parameter of the hexadecapole
deformation. Using $\lambda_4 Q_4$ as an additional constraint,
allows, indeed, to vary somewhat the neck radius of the drop.
However, at large value of $\lambda_4$ the $\lambda_4 Q_4$
constraint results in very peculiar shapes.

Another possibility to vary the neck radius is to fix the amount
of matter in the neck region by introducing the constraining function
$f_4$ of the type \bel{f4} f_4=\frac{1}{V}\int dV
\rho^2(z)\exp{\left[-\left(\frac{z-z_n}{\Delta z}\right)^2\right]}.
\end{equation}
For simplicity we assume here that $\Delta z$ has the same
meaning and value as used in Sections 3-4.

\begin{figure}[ht]
\includegraphics[width=0.45\textwidth]{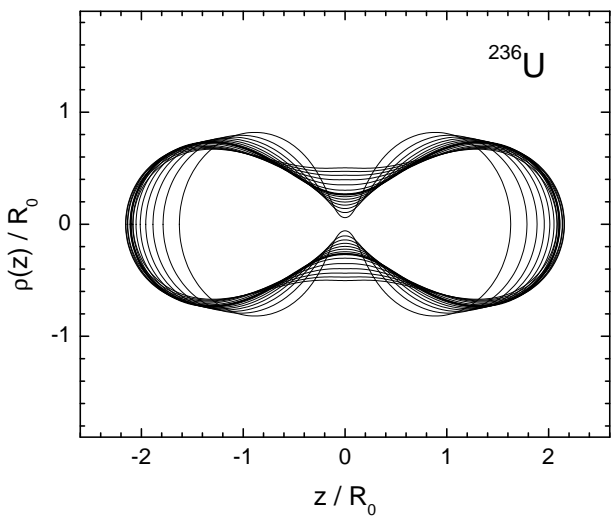}
\caption{Effect of a $\lambda_4 f_4$
restriction \protect\req{f4} on the optimal shapes: the heavy solid line shows the
profile function calculated with $\lambda_4=0$} \label{fig8a}
\end{figure}

The effect of $\lambda_4 Q_4$ on the optimal shapes is
demonstrated in Fig.~\ref{fig8a}. Indeed, varying $\lambda_4$
(keeping $\lambda_2$ fixed) allows to change the neck of the drop
in a rather broad region.
%%%%%%%%%%%%%%%%%%%%%%%%%%%%%%%%%%%%%%%%%%%%%%%%%%%%%%%%%%%%%%%%%%%%%%%%%%%%%%%%%%%%
%%%%%%%%%%%%%%%%%%%%%%%%%%%%%%%%%%%%%%%%%%%%%%%%%%%%%%%%%%%%%%%%%%%%%%%%%%%%%%%%%%%%%%%%

\begin{figure}[ht]
\includegraphics[width=0.45\textwidth]{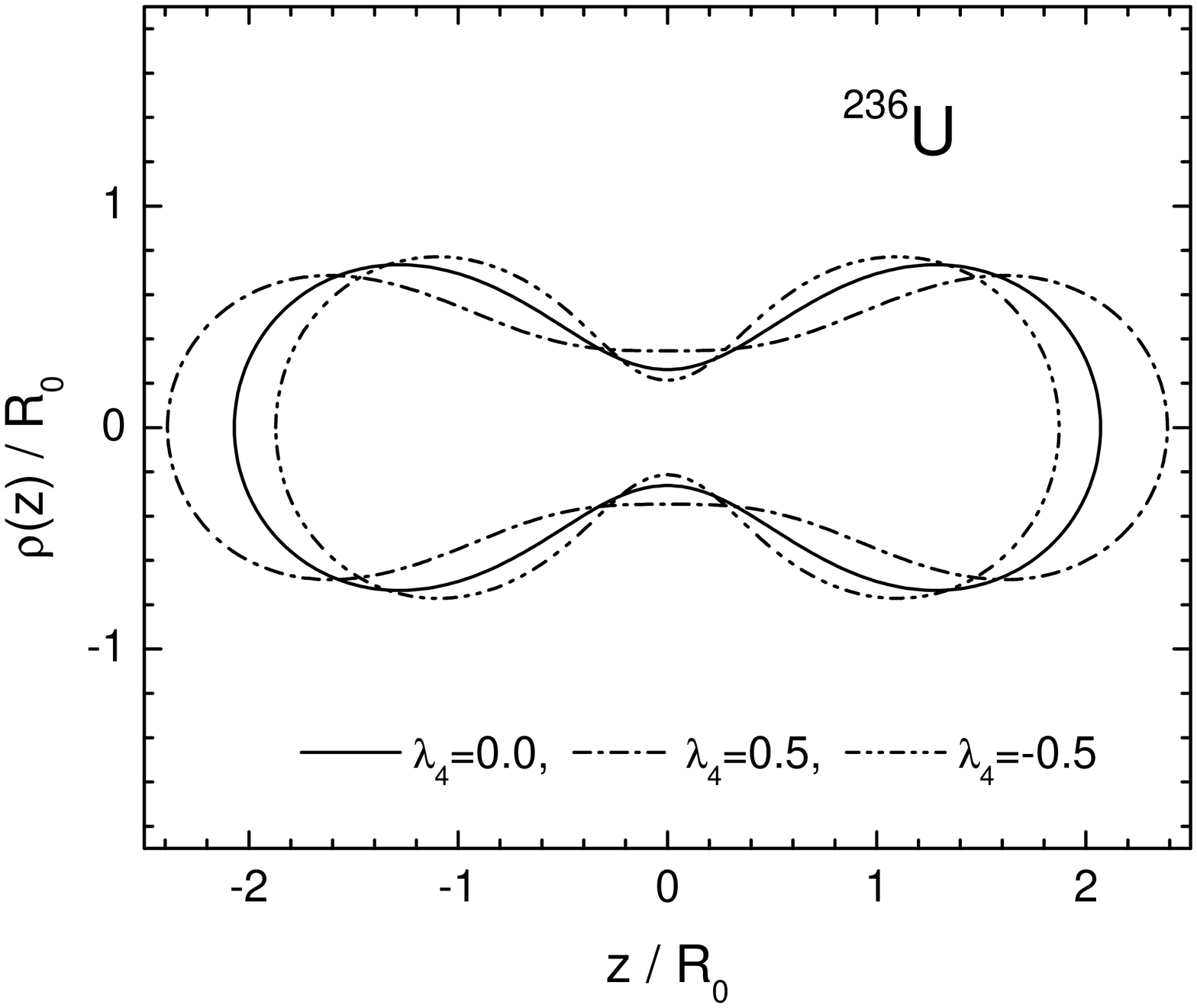}
\caption{The profile
functions $\rho(z)$ at the scission point calculated with
$\lambda_4=0$ (solid), $\lambda_4=0.5$ (dash-dot) and
$\lambda_4=-0.5$ (dash-dot-dot).} \label{fig8b}
\end{figure}

The introduction of the neck degree of
freedom has an important consequence for the scission shape.
Depending on the neck radius, the scission shapes become more
elongated or shorter, see Fig.~\ref{fig8b}. Thus, it turns out
possible to introduce the so called \cite{brosa} super-long or
super-short scission shapes which represent the possibility of the
existence
of few fission modes and are exploited by the interpretation of
the experimental data, see for example \cite{humbsch}.
%%%%%%%%%%%%%%%%%%%%%%%%%%%%%%%%%%%%%%%%%%%%%%%%%%%%%%%%%%%%%%%%%%%%%%%%%%%%%%%%%%%%
%%%%%%%%%%%%%%%%%%%%%%%%%%%%%%%%%%%%%%%%%%%%%%%%%%%%%%%%%%%%%%%%%%%%%%%%%%%%%%%%%%%%%%%%

The Coulomb interaction energy
"immediately after scission"
for the super-long or super-short shapes shown in Fig.~
\ref{fig8b} is plotted in Fig.~\ref{fig7b} by dash-dot and
dash-dot-dot lines. Qualitatively these results are very close to
the contribution from three fission modes \cite{haru} shown in
Fig.~13 of \cite{CIP}. For a more precise estimate of the
contribution from the super-long or super-short scission shapes,
full dynamical calculations (with the account of shell effects)
are required.

%%%%%%%%%%%%%%%%%%%%%%%%%%%%%%%%%%%%%%%%%%%%%%%%%%%%%%%%%%%%%%%%%%%%%%%%%%%%%%%%%%%%%%%%
\section{Summary and Conclusions}
\label{summa}
%%%%%%%%%%%%%%%%%%%%%%%%%%%%%%%%%%%%%%%%%%%%%%%%%%%%%%%%%%%%%%%%%%%%%%%%%%%%%%%%%%%%%%%%
The optimal-shape approach is put into practice by the
construction of the constraint on the mass asymmetry which is an
interpolation between the constraint on quadrupole and octupole
moments (which is quite successful at small deformations) and the
constraint on the distance between the centers of mass of the
future fission fragments and the difference of their masses
(which is well defined for the shape with a neck or separated
fragments). The use of this new constraint allows to define  the
scission point shapes in broad region of the mass asymmetries.

It is shown that the optimal-shape procedure can be further
extended by incorporating the neck degree of freedom. The
introduction of the neck degree of freedom leads to the
fission valleys, the existence of which follows from the analysis
of experimental data.

The account of
shell effects on the potential energy surface of the optimal
drops will be the subject of future studies.

\section*{Acknowledgements}
The author appreciates very much the fruitful discussions with
Profs. J. Bartel, N. Carjan, H.-J. Krappe, V.V.Pash\-kevich, K. Pomorski and is
grateful to the theory group of CENBG for the warm hospitality
during his stay at Bordeaux.

\end{document}